\documentclass[12pt]{article}

\usepackage {epsfig,subeqn}

\def\bib{\bibitem}
\def\be{\begin{equation}}
\def\ee{\end{equation}}
\def\barr{\begin{array}}
\def\earr{\end{array}}
\def\dis{\displaystyle}

\def\etal{ {\em et al.}}
\def\ie{ {\em i.e.}}

\def\lsim{\:\raisebox{-0.5ex}{$\stackrel{\textstyle<}{\sim}$}\:}
\def\gsim{\:\raisebox{-0.5ex}{$\stackrel{\textstyle>}{\sim}$}\:}
\def\mev{\: {\rm MeV} }
\def\gev{\: {\rm GeV} }
\def\tev{\: {\rm TeV} }
\def\pb{\: {\rm pb}}

\def\ra{\rightarrow}

\def\rp{$R_p \hspace{-1em}/\;\:$}

\def\ib#1,#2,#3{       {\em ibid.\/ }{\bf #1} (19#2) #3}
\def\ap#1,#2,#3{       {\em Ann.~Phys.~(NY)\/ }{\bf #1} (19#2) #3}
\def\ijmp#1,#2,#3{     {\em Int.~J.~Mod.~Phys.\/ } {\bf A#1} (19#2) #3}
\def\mpl#1,#2,#3 {     {\em Mod.~Phys.~Lett.\/ } {\bf A#1} (19#2) #3}
\def\np#1,#2,#3{       {\em Nucl.~Phys.\/ }{\bf B#1} (19#2) #3}
\def\npps#1,#2,#3{     {\em Nucl.~Phys.~B (Proc.~Suppl.)\/ }{\bf B#1}
                             (19#2) #3}
\def\plb#1,#2,#3{      {\em Phys.~Lett.\/ }{\bf B#1} (19#2) #3}
\def\pr#1,#2,#3{       {\em Phys.~Rev.\/ }{\bf #1} (19#2) #3}
\def\prd#1,#2,#3{      {\em Phys.~Rev.\/ }{\bf D#1} (19#2) #3}
\def\prep#1,#2,#3{     {\em Phys.~Rep.\/ }{\bf #1} (19#2) #3}
\def\prl#1,#2,#3{      {\em Phys.~Rev.~Lett.\/ }{\bf #1} (19#2) #3}
\def\prog#1,#2,#3{     {\em Prog.~Theor.~Phys.\/ }{\bf #1} (19#2) #3}
\def\rmp#1,#2,#3{      {\em Rev.~Mod.~Phys.\/ }{\bf #1} (19#2) #3}
\def\sp#1,#2,#3{       {\em Sov.~Phys.-Usp.\/ }{\bf #1} (19#2) #3}
\def\zpc#1,#2,#3{      {\em Z. Phys.\/ }{\bf C#1} (19#2) #3}
\def\appb#1,#2,#3{     {\em Acta Phys.\ Polon.\/ }{\bf B#1} (19#2) #3}

\begin{document}

\setcounter{page}{0}
\thispagestyle{empty}
\renewcommand{\thefootnote}{\fnsymbol{footnote}}
\begin{flushright}
CERN-TH/97-26\\[2ex]
{\large \tt hep-ph/9702392} \\
\end{flushright}
\vskip 45pt
\begin{center}
{\Large{\bf {\boldmath $R$}-Parity Violation at HERA?}} \\[2cm]
{\bf
    Debajyoti Choudhury\footnote{debchou@mail.cern.ch}
        {\rm and}
    Sreerup Raychaudhuri\footnote{sreerup@mail.cern.ch}
}

\rm
\vspace{13pt}
{\em Theory Division, CERN, CH 1211 Geneva 23, Switzerland.}

\vspace{50pt}
{\bf Abstract}
\end{center}

\begin{quotation}
We examine the possibility that the high-$Q^2$ events seen at HERA are 
due to the production and decay of squarks of $R$-parity-violating 
supersymmetry. The relevant $R$-parity-violating coupling(s) is (are)
identified and shown to lie between 0.03 and 0.26.  Consequences of such 
a coupling at other experiments, such as the LEP and the Tevatron, are 
discussed. 
\end{quotation}
\vspace{15ex}
\centerline{\it To appear in {\em Physics Letters} {\bf B}}
\vspace{15ex}
\noindent
CERN-TH/97-26\\
February 1997
\vfill
\newpage
\setcounter{footnote}{0}
\renewcommand{\thefootnote}{\arabic{footnote}}
\setcounter{page}{1}
\pagestyle{plain}
\advance \parskip by 10pt
\pagenumbering{arabic}

Recent reports of excess high-$Q^2$ events in both the H1~\cite{H1} and 
ZEUS~\cite{ZEUS} detectors at HERA are of considerable interest, since 
they correspond to possible leptoquark sightings.  
The H1 Collaboration \cite{H1} has reported 12 neutral current (NC)
events with $Q^2 > 15000$ GeV$^2$ for an
integrated luminosity of 14.2 pb$^{-1}$ against a Standard Model (SM)
deep inelastic scattering (DIS)
background of $4.7 \pm 0.76$ events. The  fluctuation probability 
for this is $9 \times 10^{-3}$.
The ZEUS Collaboration \cite{ZEUS} has reported 5 NC 
events with $Q^2 > 20000$ GeV$^2$ for an
integrated luminosity of 20.1 pb$^{-1}$ against a Standard Model 
background of $0.9 \pm 0.08$ events, for which there is a fluctuation 
probability of $1.4 \times 10^{-2}$.
The excess events appear to indicate a resonance of mass around 200 GeV.
A confirmation 
would point emphatically to new physics  
since leptoquark fields ---  vector or scalar --- are not present in the 
SM. Vector leptoquarks 
arise naturally in Grand Unified Theories (GUTs)~\cite{GUT}, but
these are constrained by the proton lifetime to be very heavy --- 
far beyond the reach of HERA energies. Relatively light vector 
leptoquarks may still exist in certain models, but these are not 
readily amenable to coupling constant unification~\cite{Sarkar}. 
On the other hand, scalar leptoquarks --- arising from Higgs 
multiplets --- are generically free from such constraints. Most such 
models, however, suffer from a major conceptual drawback, namely the
lack of a satisfactory explanation for the large mass hierarchies in 
the theory. The presence of a light leptoquark accessible at HERA
merely aggravates this problem. In contrast, the best known answer to 
the problem of mass hierarchies is provided by 
supersymmetry~\cite{mssm}, which is also known to
facilitate coupling constant unification~\cite{Amaldi}. It seems, 
therefore, that one will eventually 
require supersymmetry as well as grand unification in order to 
have a satisfactory theory involving the presence of elementary 
leptoquarks. However, 
before looking for an explanation of the HERA events within the 
framework of any supersymmetric GUT, it is interesting to ask if these 
events can be accommodated within the minimal supersymmetric extension 
of the Standard Model (MSSM)~\cite{mssm}, which is a simpler 
construction.

In its usual formulation, the MSSM respects the global conservation 
laws for the baryon number $B$ and lepton number $L$, which are valid 
in the SM. More conventionally represented as conservation of $R$-parity,
$R_p \equiv (-1)^{3B+L+2 S}$, where $S$ is the intrinsic spin~\cite{rpardef},
this feature immediately precludes leptoquark-like interactions. However, 
this symmetry was originally imposed {\em ad hoc} 
and has no compelling theoretical motivation.
It is thus important to investigate the 
consequences  of possible $R_p$-violating terms~\cite{rpar,rpar2} in 
the Lagrangian. This assumes particular significance 
in view of the fact that it would allow the squarks to have leptoquark-like 
interactions. A broken $R$-parity (\rp) would also imply that the lightest 
supersymmetric particle (LSP) is no longer stable: search strategies and 
existing bounds on MSSM masses and couplings need to be modified accordingly.

While it is 
possible to have both $L$-violating and $B$-violating \rp\ terms, 
non-observation of proton decay imposes very stringent conditions
on their simultaneous presence~\cite{SmVi_96}.
Assuming that the baryon-number-violating terms are identically 
zero~\cite{IbRo_92} helps evade this constraint in a natural 
way, apart from rendering simpler the cosmological requirement of the 
survival of GUT baryogenesis through to the present day. This survival 
can then be assured if at least one of the lepton numbers $L_i$ is 
conserved over cosmological time scales~\cite{DrRo_93}.

We thus concentrate on the $L$-violating part of the superpotential, 
which can be written (with the MSSM superfields in usual notation) as
\be
{\cal W} = \frac{1}{2} \lambda_{ijk} L_i L_j \overline{E_k}
  + \lambda'_{ijk} L_i Q_j \overline{D_k},
     \label{superpot}
\ee
where  $\lambda_{ijk} = -\lambda_{jik}$, with 
$i,j,k$ being family indices. Since we are interested 
in processes involving quarks, we shall further restrict ourselves 
to the $\lambda'$-type couplings only. The absence of $\lambda_{ijk}$
couplings can be assured by the imposition of suitable discrete symmetries
\cite{IbRo_92}. Neglecting quark mixing 
effects, the interaction Lagrangian can then be expressed as 
\be
\barr{rcl}
{\cal L}_{\lambda'} & = & \dis
- \lambda'_{ijk}
      \left[  \tilde \nu_{iL} \overline{d_{kR}} d_{jL}
            + \tilde d_{jL} \overline{d_{kR}} \nu_{iL}
            + {\tilde d}_{kR}^\star \overline{(\nu_{iL})^c} d_{jL}
           \right.  \\[1.5ex]
      & & \dis \hspace*{3.2em}
      \left.
            - \tilde e_{iL} \overline{d_{kR}} u_{jL}
            - \tilde u_{jL} \overline{d_{kR}} e_{iL}
            - {\tilde d}^\star_{kR} \overline{(e_{iL})^c} u_{jL}
      \right]
       + {\rm h.c.}
\earr
          \label{lagr}
\ee
Clearly, while the sleptons and sneutrinos behave rather like charged and 
neutral Higgses in a non-minimal model, the squark coupling mimics that 
of leptoquarks.

Given the above couplings, the processes of interest at HERA 
are~\cite{Hewett,Kon,H1RpV}
\subequations
\be
      e^+  +  d_k  \longrightarrow  \tilde u_{jL} 
                    \longrightarrow  e^+  +  d_k  
       \label{signal:usq}
\ee
\be
      e^+  +  \bar u_j   \longrightarrow  \overline{\tilde d_{kR} }
                    \longrightarrow  e^+  +  \bar u_j \ , \\
       \label{signal:dsq}
\ee      
\endsubequations
and, of the 27 possible $\lambda'$s, only nine ($\lambda'_{1jk}$)
are relevant to our discussion. If one 
admits the possibility that more than one of these couplings may be 
non-zero, {\em tree-level} flavour-changing NCs are 
introduced and the bounds on most of the {\em products} are very
severe~\cite{Prod_coup}. More conservative bounds are obtained under 
the simplifying assumption that only one of these couplings may be 
non-zero. The relevant $2\sigma$ bounds (with their sources) are :
\be
\barr{rclcl}
\lambda'_{11k} & \lsim & 
            \dis 0.03 \; ( m_{\tilde d_{kR} } / 100 \gev)
             & \quad \quad & {\rm Charged \ current\ 
                          universality}~\cite{BGH_89},
             \\[1ex]
\lambda'_{111} & \lsim & 
            \dis 10^{-4}\; 
                 m_{\tilde q} m_{\tilde g}^{1/2}  / (100 \gev)^{3/2}
             & \quad \quad & {\rm Neutrinoless\ double \ 
                               \beta{\mbox -}decay}~\cite{HKK_96},
             \\[1ex]
\lambda'_{12k} & \lsim & 
            \dis 0.4 \; ( m_{\tilde d_{kR}}  / 100 \gev)
             & \quad \quad & {\rm D{\mbox -}decays}~\cite{BhCh_95},
             \\[1ex]
\lambda'_{122} & \lsim & 
            \dis 0.03 \; ( m_{\tilde s}  / 100 \gev)
             & \quad \quad & {\rm \nu_e{\mbox -}Majorana\ 
                              mass}~\cite{GRT_93},
             \\[1ex]
\lambda'_{13k} & \lsim & 
            \dis 0.63 \; ( m_{\tilde d_{kR}}  / 100 \gev)
             & \quad \quad & {\rm D{\mbox -}decays}~\cite{BES_95},
             \\[1ex]
\lambda'_{133} & \lsim & 
            \dis 0.03 \; ( m_{\tilde b}  / 100 \gev)
             & \quad \quad & {\rm \nu_e{\mbox -}Majorana\ 
                              mass}~\cite{GRT_93},
             \\[1ex]
\lambda'_{1j1} & \lsim & 
            \dis 0.26 \; ( m_{\tilde q_{jL} } / 100 \gev)
             & (1 \sigma) & 
                    {\rm Atomic\ parity \ violation}~\cite{BGH_89}.
             \\[1ex]
\earr
         \label{constraints}
\ee 
Now, of the nine couplings $\lambda'_{ijk}$, 
only for three ($\lambda'_{1j1}$)
can a valence quark in the proton participate in the production 
mechanism (eq.(\ref{signal:usq})). 
For all other couplings, a sea-quark 
must appear in the initial state. As the data~\cite{H1,ZEUS} seem
to indicate a moderately large value for the parton momentum 
fraction, the corresponding sea-quark densities are rather 
small. An explanation of the data  
would then need a large value of the corresponding $\lambda'$, which is
already ruled out by the constraints of eq.(\ref{constraints}). 

Of the three that are left, $\lambda'_{111}$ has already been 
constrained to be too small to be relevant. Consequently, we are left 
with two possible couplings and hence scenarios, namely :
\subequations
\be
    \lambda'_{121} \quad :  \quad  
             e^+  +  d   \longrightarrow   \tilde c_{L} 
                         \longrightarrow   e^+  +  d  
        \label{scharm}
\ee
and
\be
    \lambda'_{131} \quad :  \quad  
             e^+  +  d   \longrightarrow   \tilde t_{L} 
             \longrightarrow   e^+  +  d ,
             \label{stop}
\ee
\endsubequations
so that the HERA NC events could be signals for the production of a 
left-handed scalar charm or stop. The major difference between these 
signals and those for a leptoquark (of the same mass and quantum 
numbers) lies in the fact that these squarks can have other
($R_p$-conserving) decay modes into quarks and charginos/neutralinos.
This would show up either through the branching ratio or the width 
of the resonance~\cite{BuDr_93}.

Until now, we have considered only the $T_3 = \frac{1}{2}$ component of 
the squark doublet. However, the masses of the two isospin 
components are related, with the splitting determined by the 
corresponding quark masses and the $SU(2)_L$-breaking $D$-term~\cite{mssm}:
\[
        m^2_{\tilde b_L} = m^2_{\tilde t_L} - m_t^2 + m_b^2 
                         - m_W^2 \cos 2 \beta    \ ,
\]
and similarly for the ($\tilde c_L, \tilde s_L$) pair. 
This implies the presence of at least one more relatively low-lying squark 
state ($\tilde s_L$ is somewhat heavier than $\tilde c_L$ while $\tilde b_L$
is somewhat lighter than $\tilde t_L$). 
A possible consequence is an extra contribution~\cite{DH_90} to 
$\delta \rho \equiv 1 - m_W^2/ m_Z^2 \cos^2 \theta_W$. It is easy to 
check that the contribution due to ($\tilde c_L, \tilde s_L$) is 
much smaller than the experimental errors~\cite{PDG_96}. For the 
stop solution, though, this constraint is non-trivial, especially for 
lighter stop masses within the (180--220) GeV range. As an example, for 
$m_{\tilde t_L} = 180 \gev$ and $m_t = 170 \gev$, one 
needs\footnote{This applies for the case of vanishing left-right 
squark mixing. In the presence of mixing, the constraint may be
considerably relaxed.}
$\tan \beta \gsim 4$ for consistency with $\delta \rho$ 
at the $1\sigma$ level. This also implies that the $\tilde b_L$  would be 
just beyond the energy range of LEP2.

As far as HERA is concerned, the two processes 
(\ref{scharm} \& \ref{stop}) are almost identical, the only major 
difference 
being the fact that the neutralino decay modes of the stop are 
greatly suppressed by phase-space considerations. 
Consequently, for most of our analysis, we shall
consider (\ref{scharm}) to be the representative process and will 
highlight differences only where they are crucial.

While the production cross section for squarks is obviously
quadratic in $\lambda'$,  the subsequent decay can be more complicated.
If $R$-parity is violated, the squark may indeed 
be the LSP without contradicting any 
phenomenological bounds. In this case, it would be virtually
indistinguishable from a leptoquark coupling to left-handed electrons 
and the right-handed $d$-quark. 
However, we do not make any such assumption and allow the squark 
($\tilde q_L = \tilde c_L$ or $\tilde t_L$) to have $R_p$-conserving decays :
\be
\barr{rcl} 
      \tilde q_L & \longrightarrow & q + \tilde \chi_j^0 \ , \\[1.5ex]
      \tilde q_L & \longrightarrow & q' + \tilde \chi_j^\pm \ .\\
\earr
        \label{r-cons decays}
\ee
The various branching fractions depend on both 
the size of $\lambda'$ and the quark-squark-gaugino coupling as well
as on the gaugino masses. Under 
the assumption of gaugino mass unification in the electroweak 
sector, the latter are determined 
in terms of three parameters: the gaugino mass term $M_2$ in the $SU(2)_L$
sector, the higgsino mixing parameter $\mu$ and $\tan \beta$, the ratio 
of the Higgs vacuum expectation values. Any explanation of the HERA
events within the framework of $R_p$-violating supersymmetry must, 
therefore, consider the constraints that can already be imposed on the 
parameter space from LEP data. Now, if $R$-parity is violated, most of 
the usual bounds do not hold any longer unless further assumptions are 
made. In fact, with our choice of a non-zero $\lambda'$, the only 
relevant constraint (in the $M_2$--$\mu$ plane) 
is the one derived from the {\em total width} of the 
$Z$~\cite{LEPEWG}:
\be
      \sum_{i,j=1}^2 \Gamma( Z \ra \tilde \chi_i^+ \tilde \chi_j^-) 
          + \sum_{i,j=1}^4 \Gamma( Z \ra \tilde \chi_i^0 \tilde \chi_j^0) 
               \lsim 23.1 \mev \ .
          \label{lep_constr}
\ee

\begin{figure}[htb]
\vskip 3.8in
      \relax\noindent\hskip -1.0in\relax{\includegraphics{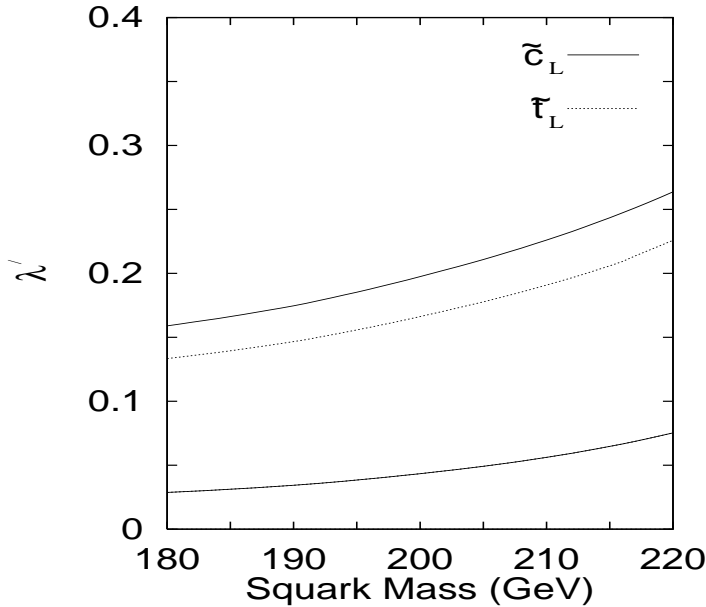}} 
\vspace{-11.5ex}
\caption{\em The range in $\lambda'$ that can account for the HERA 
   events as a function of the squark mass. The minimum value is 
   common to the two pairs $(\tilde c_L, \lambda'_{121})$ 
   and $(\tilde t_L, \lambda'_{131})$. The spread accounts for all 
   values of ($M_2, \mu, \tan \beta$) 
   consistent with the LEP constraint (\protect\ref{lep_constr}).}
              \label{fig:mass_coup}
\end{figure}
We make a simple parton-level Monte Carlo simulation of the process
$e^+ p \rightarrow \tilde c_L / \tilde t_L  + X \rightarrow e^+ d + X$
at HERA energies using acceptance cuts and selection criteria closely
modelled on those adopted for the NC signal 
by the H1 Collaboration \cite{H1}. Like
them, we use the MRS(H)~\cite{MRSH} distributions for the parton 
densities inside the proton, although we have checked that our results are not 
very sensitive to the particular structure functions being used. 
For a given value of the squark mass,  we determine
all values of the \rp-coupling $\lambda'_{1j1}$ ($j = 2,3$), which
can yield the observed excess of $12 - (4.7 \pm 0.76) \approx 7$ events in 
14.2 pb$^{-1}$ of data \cite{H1}. A value of $\lambda'_{1j1}$ is deemed 
acceptable
if we can find a set of values of $M_2,\mu,\tan \beta$ in the ranges
$( 0 < M_2 < 1 \tev), (- 1 \tev < \mu < 1 \tev), ( 1 < \tan \beta < 50)$,
which is consistent with the LEP1 bounds and
for which the required number of events is predicted. Our results are
shown in Fig.~\ref{fig:mass_coup} where solid (dotted) lines show, for
each value of the squark mass, the upper and lower permissible values 
of $\lambda'_{1j1}$ for $j = 2\;(3)$.  The lower bounds are the same
for the two cases; this corresponds to the fact that a small coupling
requires a large branching ratio into $e^+ + d$, which is only possible
if both charginos and neutralinos are very heavy, irrespective of the
mass of the quark in the decay products. Another consequence of this
fact is that the dependence of the allowed range in coupling is minimally 
dependent on $\tan \beta$.  Thus, even if low values of $\tan \beta$ are
disallowed by the $\rho$-parameter constraint for stops, the curves
remain unchanged.  It is interesting that the
coupling cannot be smaller that 0.03 or larger than 0.26 (0.22) for
$\tilde c_{L}\; (\tilde t_{L})$ production to be the explanation of the
HERA events. This result is consistent with the earlier bounds given
by the H1 Collaboration~\cite{H1RpV} based on 2.8 pb$^{-1}$ of data.

\begin{figure}[htb]
\vskip 3.8in
      \relax\noindent\hskip -1.0in\relax{\includegraphics{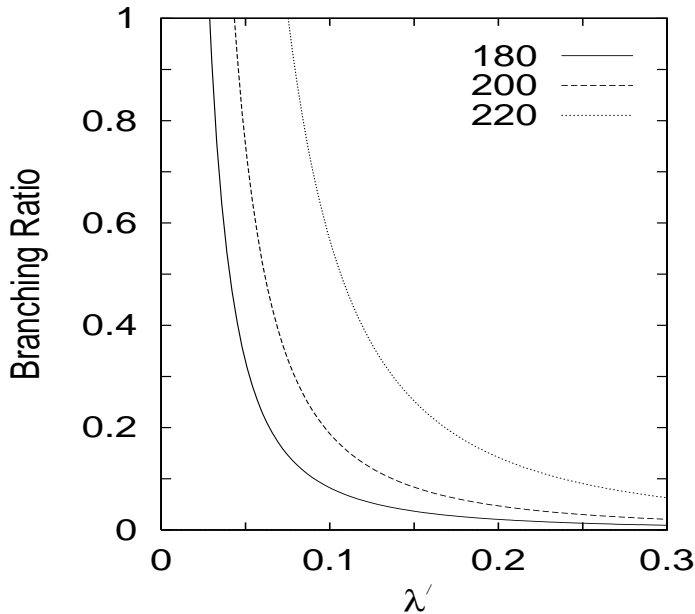}} 
\vspace{-11.5ex}
\caption{\em The $R_p$-violating branching fraction for the squark 
   as a function of the coupling for three different values 
   of the squark mass (marked, in GeV).} 
              \label{fig:coup_brfrac}
\end{figure}
In Fig.~\ref{fig:coup_brfrac}, we exhibit the branching fraction of
$\tilde u_{iL} \rightarrow e^+ + d \; (i = 2,3)$ required to explain the H1 
events for $Q^2 > 15000 \gev^2$ as a function of the relevant $\lambda'$, 
for three representative values of the 
squark mass. These curves tell us that for reasonably large values of 
the coupling the branching ratio must be rather small, \ie\ 0.25 or less
(unless the squark mass is at the lower end of the range).
This immediately points to the conclusion that there should be a large 
cross-section for the squark
in question to decay to charginos or neutralinos, which,
in turn, should be expected to have decays leading to recognizable signals. 
Thus,
under the assumption that only one $R$-parity-violating 
coupling is non-zero, the lightest neutralino will decay into 
two quarks and an electron/neutrino. For example,
\be
   \lambda'_{121} \quad : \qquad \chi_1^0 \ra c \bar{d} e^- \ , \ 
                                              s \bar{d} \nu_e
          \label{neut_decay}
\ee
as well as the conjugate modes. The situation is analogous for 
$\lambda'_{131}$, except for the fact that the decay proceeds 
only through the $ \bar{d} \chi_1^0 \tilde d$ and 
$ \bar{b} \chi_1^0 \tilde b$ couplings. Chargino decay is more 
complicated. Apart from the $R_p$-violating decay (isospin 
analogue of the process in eq.(\ref{neut_decay})), it could decay 
into two SM fermions and $\chi_1^0$ with the last-mentioned then
decaying as above. The ratio of the $R_p$-conserving and 
$R_p$-violating decay modes is determined by both the size 
of $\lambda'$ and the gaugino content of the chargino, and the 
chargino--neutralino mass splitting. These issues have 
been studied in detail in ref.~\cite{H1RpV}, for an 
integrated luminosity of $2.8 \pb^{-1}$ with the conclusion that 
such cascade decays of a squark in the (180--220) GeV range 
cannot be recognized above the background for 
$\lambda \lsim \sqrt{4 \pi \alpha}$ ($\approx 0.31$ at HERA energies). 
Since this is precisely the region of the parameter space that we are 
interested in, we may safely conclude that {\em published} constraints from 
HERA \cite{H1RpV} do not rule out any part of the parameter space shown in 
Fig.~\ref{fig:mass_coup}. On the other hand, 
{\em with the higher luminosity accumulated since then,
it may well be possible to observe such signals.} If,
indeed, such signals are seen, the signals
will be able to clearly distinguish
$R_p$-violating signals from those of leptoquarks. Non-observation
would not rule out the \rp\ option, but would restrict us to the
lowest range of $\lambda'$ for which the decay mode is
practically indistinguishable from that of a leptoquark.

\begin{figure}[htb]
\vskip 3.8in
      \relax\noindent\hskip -1.0in\relax{\includegraphics{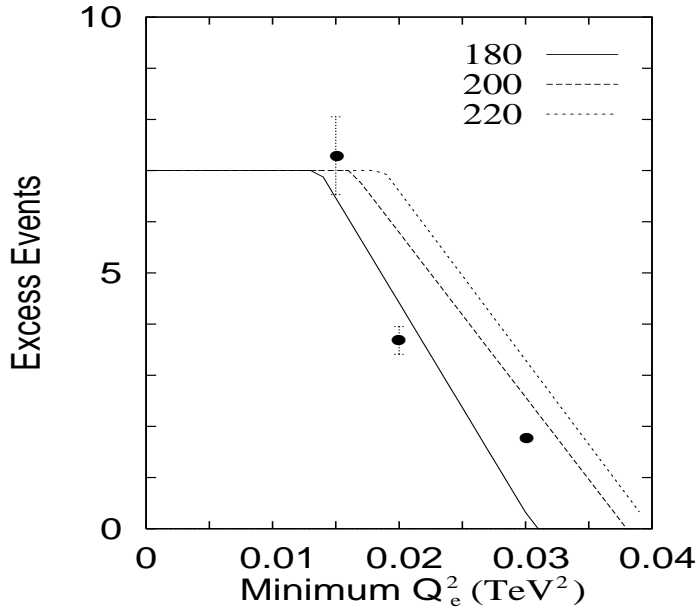}} 
\vspace{-11.5ex}
\caption{\em The number of high $Q^2$ events due to squark production as a 
       function of the minimum $Q^2$ used to bin the $e^+$, for three 
       representative squark masses (marked, in GeV). The H1 
       data~\protect\cite{H1} are also shown.}
              \label{fig:Qsquared_distrib}
\end{figure}
The hypothesis that the excess in high-$Q^2$ events is due to resonant 
production of a particle with leptoquark-like interactions is subjected
to a simple test by the data presented by the H1 Collaboration ~\cite{H1} 
for different cuts on the minimum $Q^2$. They observe an excess over DIS
predictions of approximately  7, 4 and 2 events, for 
$Q^2 > 15000, 20000, 30000 \gev^2$, respectively. This is also
consistent with the excess observed by the ZEUS Collaboration 
~\cite{ZEUS}, given their slightly higher luminosity. In 
Fig.~\ref{fig:Qsquared_distrib} we illustrate our predictions
for squark production in $R_p$-violating supersymmetry, as a function
of the cut on minimum $Q^2$, normalized to 7 events, which agrees with
 the H1 excess for $Q^2 > 15000 \gev^2$.
As before, solid, dashed and dotted lines correspond to squark masses
of 180, 200, 220 GeV. The flatness of the initial part of the curve
and its linear decrease thereafter are indicative of the relative smallness
of the $t$-channel contribution. This is as expected, since it arises
from a heavy sea-quark ($c$ or $t$) and is a non-resonant process. It is 
clear from the figure that the observations
seem to favour somewhat lower values of the squark mass within the 
range of interest.  However, for such low statistics, any
such statement must be made with a great degree of caution.

We have also examined the implications of an $R$-parity-violating solution,
for the charged current (CC) excess seen by the H1 Collaboration \cite{H1}. 
This is rather hard to explain from direct squark decays. As the neutralino 
always decays within the detector for the allowed range of $\lambda'$ 
couplings, the only possibility is to have the following process
\begin{equation}
       e^+ + p \rightarrow \tilde c_L + X \rightarrow c + \tilde \chi_i^0 + X
       \rightarrow c + s \bar s \nu(\bar \nu) + X \ ,
\end{equation}
where the neutralino decay takes place through a virtual $\tilde s_L$ squark, 
which has to be close in mass (see above) to the $\tilde c_L$. Both the
lightest and next-to-lightest neutralinos may participate in this process.
The final state 
would be purely hadronic (mostly with more than one jet) with missing momentum, 
arising from the neutrino. A parton-level Monte Carlo analysis shows that for 
the greater part of the
MSSM parameter space the final state neutrino is rather soft. In fact,
two thirds or 
more of the events are lost because of the selection cut of 50 
GeV on the missing momentum which has been imposed by the H1 Collaboration. 
For values of $\lambda'$ consistent with the NC data,
the branching ratio of the $\tilde c_L$ to neutralinos can be as high as
70\%. For the same set(s) of parameters, the  branching ratio for the
$cs \bar s \nu(\bar \nu)$ mode is around 40\%. Convoluting these with the
production cross-section, we predict up to 3 excess events in the CC sample
for  $Q^2 > 15000$ GeV$^2$, depending on the choice of the point 
in the parameter space. The excess
drops rather sharply for $Q^2 > 20000$ GeV$^2$, however, which does not seem
to agree well with the trend of the H1 data. Pushing the
comparison further is not very 
meaningful in view of the low statistics. It is also noteworthy that the 
ZEUS Collaboration \cite{ZEUS} has not reported any CC events, 
although their luminosity is somewhat higher. A full investigation must,
therefore, await more data. Meanwhile we can say with some certainty that
a charged current excess at the level observed so far can be accommodated
quite easily within our \rp-scenario.

It may be said, therefore, that the observed excess in high-$Q^2$
events at HERA can be simply explained by squark production in
$R_p$-violating supersymmetry. However, unless a further
search for chargino/neutralino decay modes of these squarks is
made and actual signals found, it will be difficult to decide whether
the present signals are those arising from leptoquarks or squarks.
We turn, therefore, to other possible consequences of the proposed 
scenario, with special emphasis on those that might distinguish 
it from the leptoquark alternative. We make the conservative 
assumption that no other supersymmetric particle would be discovered 
directly. 

We start with LEP1 results on $R_b$ 
$ \equiv \Gamma(Z \to b \bar{b}) / \Gamma(Z \to {\rm hadrons})$. 
This is of relevance only to the ($\tilde t_L$, $\lambda'_{131}$) solution.
It might be argued that a relatively light stop could contribute 
significantly to this ratio, especially if the chargino were light too.
One must remember, however, that we are dealing with $\tilde t_L$ and the 
effect is smaller compared to that for $\tilde t_R$. Explicit computations 
show that the extra contribution is of the order of $10^{-5}$ to $10^{-4}$
and hence is consistent with the observations~\cite{R_b_expt}. 
If we assume a degenerate ($\tilde t_L, \tilde t_R$) pair (this does
not affect the HERA signal), one can get a contribution as large as 
$10^{-3}$, which is also consistent with observations.
The contribution due to the $\lambda'_{131}$ coupling 
itself is small~\cite{BES_95}.
The situation for $R_\ell$ is similar~\cite{BES_95}. 
For $\lambda'_{131} \sim 0.25$ or less, the excess contribution is
always well within the experimental errors at $1\sigma$. Essentially,
contributions to both $R_b$ and $R_\ell$ are severely suppressed by a
squark mass in the 200 GeV ballpark.  Thus precision
measurements at LEP1 will not help to confirm or deny our hypothesis.

Since these squarks are too heavy to be produced at LEP2, one could only 
consider virtual effects. A relevant process is 
$e^+ e^- \ra c \bar{c}$. A non-zero $\lambda'_{121}$ leads to
an additional ($t$-channel) diagram. A study of the angular distribution 
of the charm jets may rule out the high $\lambda'_{121}$ end of the 
parameter space~\cite{DC_95}. It is only fair to point 
out, however, that this is just the range where
further searches at HERA itself may be effective.

At the Tevatron, there is enough energy to produce a 
pair of these squarks. Both $R_p$-conserving and $R_p$-violating processes 
will contribute. The cascade decay modes (through charginos
and neutralinos) have been studied in the literature~\cite{Rp_in_Tev}
in fair detail. The upshot of these is that squark masses and 
couplings of the magnitudes conceived of here would not lead to 
signals distinguishable from the background.  On the other hand,
direct decay into $e^+ d$ (in common with the leptoquark signal) will give 
rise to two hard jets and a dielectron pair. This signal might be of 
significance \cite{D0_leptoquark}, especially for low values of $\lambda'$. 
However, in view of the large SM backgrounds, a detailed and separate study 
seems to be called for \cite{selves}.

Virtual effects at the Tevatron are not likely to be a decisive 
discriminator. A non-zero $\lambda'_{131}$ will enhance the $t\bar{t}$ 
production cross-section, but the relatively large errors in the 
cross-section measurement~\cite{Top_CDF_D0} do not allow us to 
probe $\lambda'_{131} \lsim 0.3$ unless the selectron is light enough 
to be seen at LEP2~\cite{GRS_96}. The analogous process for 
$\lambda'_{121}$ would result in an enhancement of the large-$p_T$ 
dijet rates. An improvement in the angular measurement~\cite{CDF_dijet}
of the pure dijet spectrum could, in principle, rule out the large 
$\lambda'_{121}$ end of the parameter space. This, however, is unlikely 
to substantially improve upon the bounds derivable from $c \bar c$
production at LEP2. Moreover,
as pointed out already, this range of values of $\lambda'$ can probably
be probed at HERA itself. Dilepton production, on the other hand, 
can probe both the couplings~\cite{BCS_95}, but would not be able to 
distinguish between them or from the leptoquark alternative. Thus,
although they are certainly worth a detailed investigation, one cannot expect
immediate confirmation or otherwise from virtual effects.

Another interesting possibility, which arises in models with $R_p$-violation, 
is that of single squark production at the Tevatron, the process 
relevant to the present investigation being
$ g + d \ra \tilde u_{jL} + e^- \ $, 
where $j = 2,3$. With the squark decaying through the \rp\ interaction,
we are led to the signal $ p + \bar p \ra e^+ + e^- + {\rm jet} + X $.
However, as Fig. ~\ref{fig:coup_brfrac} shows, choosing a large value of
$\lambda'$ to get a sizeable cross-section here 
is more than offset by the HERA requirement of a small
branching ratio of the squark to $e^+ d$. As a result, the
signal is strongly suppressed and is not likely to be distinguishable 
above the SM background.
It may still be possible to detect signals of the chargino or neutralino
decay modes of the squark (as at HERA), especially with added 
luminosity, such as will be available with the commissioning of
the Main Injector. Eventually, perhaps, CDF/D0 will be in a position to 
confirm the signal or, at least, rule out parts of the parameter space.

In summary, then, we have investigated the possibility that the
excess high-$Q^2$ events observed at HERA are due to the production
of squarks of $R$-parity-violating supersymmetry. We have shown that
in view of low-energy constraints, this is acceptable only if one 
of the couplings $\lambda'_{121}, \lambda'_{131}$ is non-zero, so that 
the produced particle must be a left-handed scalar charm or stop.
The coupling must lie between 0.03 and 0.26 for the first option
and 0.03 and 0.22 for the second, depending on the mass of the 
squark, which, according to the data, must lie around 200 GeV
with an uncertainty of about 10\%. It turns out that, for a large
range of these couplings, large branching ratios of the produced
squark into charginos and neutralinos are predicted. Decays of the
latter should lead to distinctive
signals at HERA. On the other hand, small values of the \rp-violating
coupling would make it extremely difficult to distinguish squark
signals from those arising from leptoquarks of similar masses and
couplings.
We also show that confirmation (or otherwise) of our hypothesis at the 
two running high-energy facilities, LEP and the Tevatron, is unlikely 
to be immediately 
forthcoming, although we might expect some results from the Tevatron 
eventually with higher luminosities. 
The supersymmetric solution would thus remain a
valid and exciting hypothesis to explain the HERA events, and we look 
forward to new results in this area in the near future. 

The authors are grateful to J.~Kalinowski for drawing our attention 
to the HERA events, to M.~Drees for useful information and to 
C.E.M.~Wagner for discussions. 
The parton distributions were calculated using the package 
PDFLIB~\cite{PDFLIB}.

\end{document}